\documentclass[12pt]{article} \usepackage{epsf} 
\usepackage{graphics,graphicx}
\usepackage{subfigure}
\usepackage{ulem}

\begin{document}

\setcounter{page}{1} \makeatletter \def\@oddfoot{\hbox to \textwidth{\hfil 
\thepage\hfil}} \makeatletter \def\@evenfoot{\hbox to \textwidth{\hfil 
\thepage\hfil}}

\title{Lighweight Target Tracking Using Passive Traces in Sensor Networks
\thanks{This work was partially supported by the IST/FET/Global Computing Programme of the European Union, 
under contact number IST-2005-15964 (AEOLUS).}} 
\author{Andrei Marculescu  \thanks{ CUI, University of Geneva, Switzerland. E-mails: {\tt 
\{Andrei.Marculescu, Olivier.Powell, Jose.Rolim\}@cui.unige.ch  } }
\and Sotiris Nikoletseas \thanks{Computer Technology Institute and
Department of Computer
Engineering \& Informatics, University of Patras, Greece.
E-mail: {\tt nikole@cti.gr}. }  
\and Olivier Powell \footnotemark[2]
\and Jose Rolim \footnotemark[2] 
}

\date{\today}

\maketitle 

\begin{abstract} 

We study the important problem of tracking moving targets in wireless sensor networks. We try to overcome the limitations of standard state of the art tracking methods based on continuous location tracking, i.e. the high energy dissipation and communication overhead imposed by the active participation of sensors in the tracking process and the low scalability, especially in sparse networks. Instead, our approach uses sensors in a passive way: they just record and judiciously spread information about observed target presence in their vicinity; this information is then used by the (powerful) tracking agent to locate the target by just following the traces left at sensors. Our protocol is greedy, local, distributed, energy efficient and very successful, in the sense that (as shown by extensive simulations) the tracking agent manages to quickly locate and follow the target; also, we achieve good trade-offs between the energy dissipation and latency.        

\end{abstract}




\section{Introduction}

Recent advances in micro-electromechanical systems (MEMS) and wireless communications have enabled the development of very small,
smart, low cost 
sensing devices (\cite{intro2, intro3}) with sensing, data-processing 
and wireless transmission capabilities. They are meant to be pervasively deployed into forming ad-hoc wireless sensor networks 
that collect information from the ambient environment and make it 
available to the user. Some applications imply deployment in remote or hostile environments (battle-field, 
tsunami, earth-quake, isolated wild-life island, space exploration) to assist in tasks such as target tracking, enemy intrusion detection, forest fire detection, environmental or biological monitoring. Some 
other applications imply deployment indoors or in urban or controlled environments. Examples of such 
applications are industrial supervising, indoor micro-climate monitoring (e.g. to reduce heating cost by detecting poor building thermal insulation), smart-home applications, patient-doctor health monitoring or blind 
and impaired assisting. Sensor networks imply distributed 
and collaborative data-processing, because of the small utility of each sensor individually
and the severe resource constraints, mainly with respect to energy but also memory and computation capabilities.

\subsection{Tracking Problems and Our Approach}

We wish to solve the problem of tracking objects moving in a domain over a certain period of time i.e. we want the wireless sensor network to be able to detect the position of any moving object in the domain at any time in the monitoring period. Location tracking is the current standard approach to the tracking problem in sensor networks. Location tracking includes several phases: target detection, distance estimation, position evaluation and trajectory estimation. The target detection is performed individually by sensor nodes in the network and does not need any synchronization. The distance estimation uses either an empiric law (e.g. based on signal attenuation estimations) translating the sensing intensity into a real valued distance, or uses a special hardware device performing this translation. The position estimation phase is typically addressed with trilateration. At least three nodes are periodically chosen among the sensors located next to the target. Each of these three nodes gives an estimate of its distance to the target. Finally, another node (that can be one of the three) performs trilateration based on the distance estimates and computes the position estimate. The position evaluation is then sent to a base-station in order to estimate the trajectory of the moving object.
   
   While this method can give very precise results when the network is dense enough (at least 3 nodes have to be located next to the target as it moves), it has several drawbacks. First, this technique does not scale to the case of tracking multiple objects, for instance a group of objects moving in formation, or several individual objects. Protocols designed to compute in a precise manner the position of an individual target are not able to cope with the question ``detect any of the members of a given group''.

Second, the protocol overhead is high in terms of communication. Each time a position is estimated, three nodes are chosen. Then, among these three nodes another node is chosen to perform the position evaluation. This is done through exchange of protocol messages and can be quite energy consuming. While precision may be needed in some applications, other applications need only rough information on the location area. For instance, if the target is represented by a group of several objects, we may want to detect just an arbitrary location within this area.
 
Finally, this technique assumes the existence of a base station with powerful computational resources, generally supposed to be fixed. Although this is sometimes realistic, many of the tracking applications do not respect this hypothesis. For instance, soldiers tracking an enemy would be moving to follow it. Another example is habitat monitoring: a group of biologists tracking an antelope herd also needs to move towards the detected herd. In fact, many natural tracking applications actually use a mobile sink. 

   In order to cope with these problems we broaden the hypothesis defining the tracking problem: a) The target can be an individual object (e.g. a patient in a hospital) as well as a large group of several objects (e.g. an animal herd). b) The tracking precision becomes a parameter of the problem. We are primarily concerning ourselves with fuzzy tracking applicable to large group of objects, where the degree of precision needs not be high. But our techniques can use any sort of location information attached to sensors to enhance our solution's quality (for instance, if in a hospital sensor nodes are aware of their location in terms of building wing and room number, and we can use this information for the tracking results). c) The sink can be either fixed or mobile and there can be more than one sink (for instance, troopers or a group of biologists). As the fixed sink case is rather well studied, we are concerning ourselves mainly with the mobile sink problem, which is useful in applications but also more challenging to cope with.

\subsection{Our Contribution}

The main idea of our approach is to avoid actively involving sensors in the tracking process. Instead, we exploit ``traces'' of target presence that are anyway left around its moving trajectory. The role of sensors in our protocol is rather passive: they just locally decrease trace intensities with time (to take into account the fact that the target was detected but then moved away) and also propagate them appropriately, in order to spread then in a balanced way in the network at a low energy cost. The active role in our tracking approach is performed by the tracking agent, that greedily follows trace gradients to locate the moving target.

As shown by the simulation findings, our approach achieves significant improvements over well known methods in the state of the art. First, our protocol is successful (i.e. the target is indeed located), even in the case of multiple targets and mobile sinks. Also, our protocol is very efficient, since it reduces energy a lot (by avoiding active sensor participation) while keeping latency low (by spreading trace intensities in a way that ``covers'' the network in a balanced way).

\subsection{Related Work and Comparison}

A standard centralized approach to tracking (\cite{das}), is ``sensor specific'', in the sense
that it uses some smart powerful sensors that have high processing abilities.
In particular, this algorithm assumes that each node is aware of its absolute location (e.g. via a GPS) or of a relative location.
The sensors must be capable of estimating the distance of the target from the
sensor readings. The process of tracking a target has three distinct steps: 
detecting the presence of the target, determining the direction of motion of the target
and alerting appropriate nodes in the network.
Thus, in their approach a very large part of the network is actively involved in the tracking process,
a fact that may lead to increased energy dissipation. Also,
in contrast to our method that can simultaneously handle multiple targets, 
their protocol can only track one target in the network at any time.
Overall, their method has several strengths (reasonable estimation error, precise location of the tracked source,
real time target tracking, but there are weaknesses as well
(intensive computations, intensive radio transmissions).

Our method is entirely different to the network architecture design approach
for centralized placement/distributed tracking (see e.g. the book \cite{book} for a nice overview). 
According to that approach, optimal (or as efficient as possible) sensor deployment strategies are proposed to ensure maximum
sensing coverage with minimal number of sensors, as well as power conservation in
sensor networks. In one of the centralized methods (\cite{grid}), that focuses on deployment optimization, 
a grid manner discretization of the space is performed. Their method tries to find 
the gridpoint closest to the target,
instead of finding the exact coordinates of the target. In such a setting,
an optimized placement of sensors will guarantee that every gridpoint
in the area is covered by a unique subset of sensors. 

Another network design approach for tracking is provided in \cite{nikole}, that tries to avoid an expensive massive deployment of sensors, taking advantage of possible coverage ovelaps over space and time,
by introducing a novel combinatorial model (using set covers) that captures such overlaps. The authors then use this model to design and analyze an efficient approximate method for sensor placement and operation, that with high probability and in polynomial expected time achieves a $\Theta (\log n ) $ approximation ratio to the optimal solution. 

Clearly, in contrast to our direct approach to tracking,
such network design solutions can provide full tracking only when combined with collaborative
processing methods, to process and synthesize individual target location estimations.

As opposed to centralized processing, 
in a distributed model sensor networks distribute the computation
among sensor nodes. Each sensor unit acquires local, partial, and relatively
coarse information from its environment. The network then collaboratively determines
a fairly precise estimate based on its coverage and multiplicity of sensing
modalities. Several such distributed approaches have been proposed.
In \cite{liu}, a cluster-based distributed tracking scheme is provided. The sensor network is logically
partitioned into local collaborative groups. Each group is responsible for providing
information on a target and tracking it. Sensors that can jointly provide the
most accurate information on a target (in this case, those that are nearest to the target)
form a group. As the target moves, the local region must move with it; hence
groups are dynamic with nodes dropping out and others joining in.
It is clear that time synchronization is a major prerequisite for this approach to
work. Furthermore, this algorithm works well for merging multiple tracks corresponding to the
same target. However, if two targets come very close to each other, then the mechanism
described will be unable to distinguish between them.

Another nice distributed approach is the dynamic convoy tree-based collaboration (DCTC) framework that has been proposed
in \cite{infocom}. The convoy tree includes sensor nodes around the detected target, and the
tree progressively adapts itself to add more nodes and prune some nodes as the target
moves. In particular, as the target moves, some nodes lying
upstream of the moving path will drift farther away from the target and will be
pruned from the convoy tree. On the other hand, some free nodes lying on the projected
moving path will soon need to join the collaborative tracking.
As the tree further adapts itself according to the movement of the target, the root
will be too far away from the target, which introduces the need to relocate a new
root and reconfigure the convoy tree accordingly.
If the moving target's trail is known a priori and each node has knowledge about
the global network topology, it is possible for the tracking nodes to agree on an
optimal convoy tree structure; these are at the same time the main weaknesses of the protocol, since
in many real scenarios such assumptions are unrealistic.   

Finally, a ``mobile'' agent approach is followed in \cite{tseng}, 
i.e. a master agent is traveling through the network, and two slave agents are assigned the task to participate to the trilateration.
As opposed to our method, their approach is quite complicated,
including several sub-protocols (e.g. election protocols, 
trilateration, fusion and delivery of tracking results, maintaining a tracking history).
Although by using mobile agents, the sensing, computing and communication overheads can be greatly reduced,
their approach is not scalable in randomly scattered networks
and also for well connected irregular networks, since a big amount of offline computation is needed
Finally, the base that receives the tracking results is assumed fixed (in a tracking application this can be a problem).

The interested reader is referred to {\cite{guibas}},
the nice book by F. Zhao and L. Guibas, that even presents
the tracking problem as a ``canonical'' problem
for wireless sensor networks. Also, several tracking approaches
are presented in \cite{book}.

\section{Our Tracing Handling Tracking Protocol (THTP)}

\subsection{Protocol Overview} 

   The intuitive idea behind our tracking protocol is inspired by a natural model: the way a tiger tracks an antelope in a Savannah. Antelope traces are initially stored into the environment as the antelope moves. The intensity of these initial traces decreases with time. These traces are partially spread through the Savannah (e.g. due to the wind action) and the intensity of a spread trace decreases with the distance from the initial trace. Thus, the only role of the environment is to store and to rather passively spread traces. It represents a passive actor of the tracking process. The active actor of the tracking process is the tiger, which senses the traces stored into the environment and tries to track the antelope by following the trace gradient.
   
   The correspondence to our problem is the following: the sensor network is the environment, the tiger is a tracking agent and the antelope is the tracking object. Using this natural model has the advantage that as long as no tracking demand is issued, no (or few resources) are used into the network, as opposed to the location tracking algorithms where the network is continuously tracking the object in a proactive manner, hence consuming valuable energy. A passive network is the key to our energy optimisation. On the other hand, when the tracking is (reactively) in progress, only local computations and very light computation resources are needed in order to follow the trace gradient.
   
   When the tracking is needed, a tracking agent starts to walk through the network. This agent follows the trace gradient in a greedy manner. The agent can be either a pure software agent originated by a mobile sink, or a human provided with a special device communicating with the network, or even a mobile robot interacting directly with the network. The case of a software agent is somehow more complicated, because it has to send the results back to the originator of the agent, typically a mobile sink. That is the reason why we focus on the case of a software agent in the present work.

\subsection{Detailed Protocol Description}

\subsubsection{Trace storage, spreading and attenuation.}
\label{trace spreading}

Each trace can be represented by the following record:

\begin{table}
\centering
 \begin{tabular}{|c|l|}
  \hline
  \textbf{Attributes} & \textbf{Possible values}\\\hline\hline
  \texttt{TYPE} & \texttt{INITIAL} or \texttt{SPREAD}.\\\hline
  \texttt{START\_TIME} & Time at which the trace was initialized.\\\hline
  \texttt{START\_INTENSITY} & Initial intensity of the trace.\\\hline
  \texttt{INTENSITY} & A positive real number smaller than $300$.\\\hline
  \texttt{PATH} & A list of $1$ to $3$ parent traces.\\\hline
  \texttt{MAX\_INTENSITY} & $300$ \\\hline
  \texttt{TARGET\_ID} & Used for the tracking multiple targets.\\\hline
\end{tabular}
\caption{Attributes of traces}
\end{table}


   When a sensor node detects a target, it stores a trace with the given \texttt{TARGET\_ID} (or group ID or both, depending on the target being an individual object or a group), maximum intensity (in our simulations, $\texttt{MAX\_INTENSITY} = 300$) and of type \texttt{INITIAL}. If a trace of the corresponding ID already exists, its intensity is set to \texttt{MAX\_ INTENSITY}. That means that a sensor containing a trace with \texttt{MAX\_ INTENSITY} is assumed by the protocol to be next to the target. After having stored the trace, it spreads it according to the spreading strategy. 

\begin{figure}[hbt]
\centering
\includegraphics[width=0.6\textwidth]{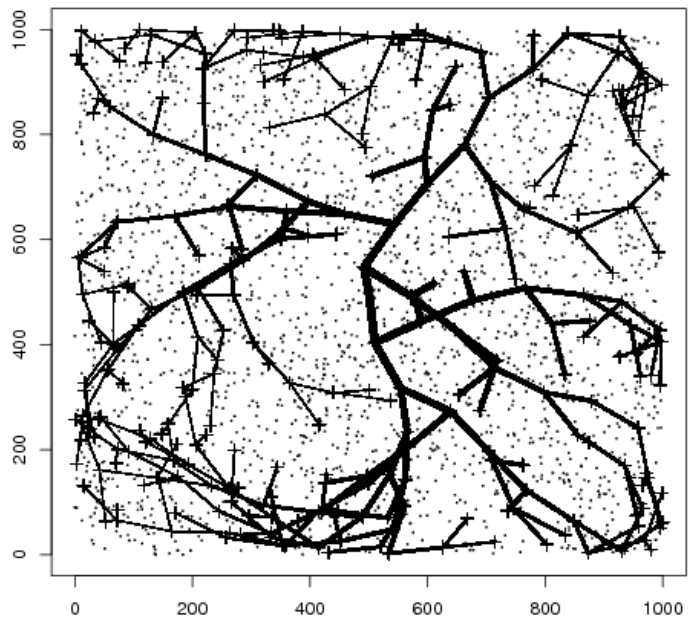}
\label{tree fig}
\caption{The efficient spanning of a trace starting at the center.}
\end{figure}

   We propose a spreading strategy which aims at covering a large area of the network, with a small amount of energy consumption. We try to build a tree of degree 2 across the network, but in a distributed manner. The node that detects the target sends a spreading message to two of its neighbours. If the receiving node has no trace of ID specified in the message it stores locally the trace and forwards the message to two of its neighbours, and so on. These two neighbours are selected according to a subtle heuristic that was designed to span the network efficiently. Suppose the node $n_0$ needs to spread its trace to two neighbours $n_1$ and $n_2$. The first neighbour, $n_1$, is selected randomly among the subset $\mathcal{V}_{rep}( n_0 )$ of neighbours of $n_0$ farther from a \textsl{repulsion point} than $n_0$. While the second node is also selected in $\mathcal{V}_{rep}(n_0)$, it is not selected randomly but deterministically as being the further away from the  point. At the end of the process, we obtain a tree of degree two spanning the network. This process is illustrated in figure \ref{tree fig}, where a trace is being propagated from the center of a network of $2500$ nodes randomly and uniformly distributed in a $1000\times 1000$ square and the communication radius of the nodes is $100$ meters. The rationale behind this heuristic is that the random choice for $n_1$ will help ensuring that the tree does not leave any uncovered holes inside the global covered region, while the idea of choosing $n_2$ according to the  point heuristic is to ensure that the tree indeed spans through the whole network. Of course this implies choosing wisely the repulsion point. Also, even though some overlapping of the branches of the tree is possible because of the random nature of the tree construction, this unpleasant possibility is reduced by introducing an inhibition mechanism that permits to limit the propagation of traces that would induce a branch overlap. As a consequence, only a small amount of messages is required to span the network since the tree branches do not overlap. 
   \paragraph{The repulsion point}
   In order to insure that the tree spreads the whole network, traces carry with them a \texttt{PATH} variable. When a trace is of type \texttt{INITIAL}, the path is empty.
   When a trace is of type \texttt{SPREAD} the path is not empty. In fact, only the last two
   hops of a path need to be kept in the \texttt{PATH} variable. When a node $n_0$ tries to spread a trace, it sets it's repulsion point to be its grand-parent in the tree rooted where the \texttt{INITIAL} trace was created. If the \texttt{PATH} is not of size two but of size $1$, it sets the repulsion point to be its parent in the tree. Finally, when a trace is of type \texttt{INITIAL}, the repulsion point is the node $n_0$ itself. As a consequence, the node $n_0$ tends to choose $n_2$ in a way that preserves some kind of inertia with respect to the two previous hops: the trace tends to go far away from where it comes. This is balanced by the fact that the other node, $n_1$, is chosen randomly.

\paragraph{The Inhibition Mechanism}
The \texttt{INITIAL\_INTENSITY} if a trace of type \texttt{INITIAL} is initialised to \texttt{MAX\_INTENSITY}.
Furthermore, we assume that each node has a clock, and that whenever a trace is being added to a node, the \texttt{INITIAL\_TIME} at which the trace has been started is recorded. Please note that we do not require the clocks of different nodes to be synchronized. At any given time $t$, the \texttt{INTENSITY} of a trace is defined to be $\max\left\{ 0, \texttt{INITIAL\_INTENSITY}  - (t - \texttt{INITIAL\_TIME}) \right\}$. When a node $n$ ``spreads a trace'' $T$ to another node $n'$, what actually happens is that a new trace $T'$ is initialised on the receiving node $n'$. The \texttt{PATH} of $T'$ is updated by appending $n$ to it. If the spreading occurs at time $t$,
the \texttt{INITIAL\_TIME} of $T'$ is set to $t$, while the \texttt{MAX\_INTENSITY} of $T'$ is set to be the \texttt{INTENSITY} of $T$ at time $t$ minus a ``spreading penalty''. In our experiment, the penalty was set to be $1$. The purpose of the spreading penalty is to ensure that a tracking agent following a gradient of ever more intense traces will actually end up at a trace of type \texttt{INITIAL}. The spreading inhibition mechanism is the following. First of all, evidently, a trace is only spread if its intensity is greater than $0$. Second,
if a node $n$ is about to spread a trace of intensity \texttt{INTENSITY} but that there is at least one member of $\mathcal{V}_{rep}(n)$, the spreading is inhibited. This is how branch overlapping in the tree is reduced.

   Furthermore, each node continuously executes a (very simple and local) intensity attenuation process. We used for our experiments a linear attenuation: at each step, the trace intensity is decremented by one. If the trace reaches the minimum value, it is deleted. The attenuation frequency (time between two consecutive steps of this algorithm) has to be lower than the time needed by the target to cover the distance of one hop. If this condition is fulfilled, the assumption that a �node with maximum trace intensity is on track� always holds. We note that other attenuation functions 
(e.g. faster) can be used in our protocol to better fit different scenarios and achieve different trade-offs.

\subsubsection{Tracking process.}

   We designed an agent-oriented tracking. When the sink (e.g. a biologist) wishes to discover the target (e.g. an antelope), it initiates a tracking agent. The tracking agent follows the trace gradient in a greedy manner. This simple process only requires for each node to be aware of the intensities of its neighbours. The decision of the agent to go to a node or to another is guided by a purely distributed criterion. During the initialisation phase, while the agent finds itself into a node without any trace, it walks randomly to any of the neighbours. Although the process is simple and distributed, there are two points we should be aware of. 
   
   First of all, the tracking applications generally need a reporting protocol, with reasonable response delay (latency). The random walk from the initialisation stage can be quite inefficient. That is why a good coverage of the spreading technique is vital for the performance of the algorithm. The better is the trace coverage the higher is the probability of a random walk to cross a trace path and then follow it. On the other hand we can use a biased random walk in order to decrease the initialisation time. The simplest is the following. Nodes are recording a special trace for the tracking agent(s). There is no spreading for these tracks, they only attenuate and vanish with time. As opposed to the antelope tracks, these tracks are inhibitory for agents. This can be viewed as a partial self-avoiding random walk.
   
   The second problem comes from the hill climbing technique used. As any other greedy technique, our trace following is subject to local minimum traps. To fix this problem, a special type of inhibitory trace is used. When the tracking agent comes to a node with the largest local intensity, it has to move back. Before moving back, the agent marks this node as being bad. Because as long as the trace gradient doesn't change such a maximum is a cul-de-sac, and the node is then avoided by this agent and any other agents. Care must be taken if several types of targets are tracked. In this case, a distinct ``bad node'' trace should be used for each type of target.
   
   In order to optimize the tracking process, several agents can be sent through the network.
   
   Let us also note that in the classical case of a fixed base station and continuous monitoring of the target, the tracking agent can be initiated by the first node having detected the target. Sending back the results is an easy process in this case. If trajectory estimation is needed, three agents are initiated instead of one.

\subsubsection{Sending back the results} 

   This part of the protocol boils down to routing a message to a mobile destination (e.g. biologist). The routing nodes are not aware of the current position of the destination. We call this type of routing pervasive routing, because its results have to be available ``everywhere in the network''.
   
   We suggest two possible strategies for solving this problem. 
   
   The first strategy is to consider the pervasive routing as an inverted tracking problem. As our protocol scales well for several targets, we can view the tracking data destination as such an object. The return message tracks its destination exactly in the same manner a tracking agent would track its target. This technique can be very efficient when the moving pattern of the destination is very dynamic (i.e. the destination covers a large area).
   The second strategy is a hybrid between the geographic routing and the tracking problem. The tracking agent knows its initial position, so when it is sending a return message, it is routed with geographic routing to the initial position, and from the initial position the destination is tracked. This technique can be very efficient when the moving pattern of the destination is almost stationary (i.e. the destination cover a very small area).

\section{Experiments}

For simulation purposes, we take the following parameters.
The network is composed of $300$ nodes. The communication radius is $100$ meters, the detection radius is $25$ meters,
the target speed is $6$ km/h. The message transmission frequency is of $1$ message per second (this determines the speed at which
traces can spread through the network, as well as the delay between two hops of the tracking agent). The density of the network is
the number of sensors per square meter. Since the number of sensors is fixed (300), the nodes are spread in a square region with
side $l$ such that $300/l^2 = density$.
While the target moves inside the network, nodes detect it. At any given time, the last which
has detected the target is the one with the highest intensity trace. Call this node $n_{best}$. At the same time, the tracker agent visits nodes. At any given time, we can consider the node with highest intensity it has visited so far. Call this node $b_{bestEstimation}$. In our simulations, we measure the distance between $n_{best}$ and $n_{bestEstimation}$. We consider
that the tracker's best estimation of the target's localization is \emph{correct}, i.e.
the target is \emph{localized} by the tracker, whenever $n_{best} = n_{bestEstimation}$.
For simulation purposes, we let some parameters change in the following way.
The results of our simulations when simulating an execution time of $20$ minutes are presented below.

\begin{table}
\centering
\begin{tabular}{|c|c|l|}
\hline 
\textbf{semantic} & \textbf{symbol} & \textbf{default value} \\\hline\hline
number of nodes & $n$ & 300 sensors\\\hline
communication radius & $d_{trx}$ & $100$ meters \\\hline
sensing radius& $d_{dtx}$ & $25$ meters \\\hline
target speed & $detection\_radius$ & $6$ km/h \\\hline
message propagation frequency& $freq$ & $1$ MSG/second \\\hline
network density& $dens$ & $10 / (100^2\cdot 3.14)$* Mag/second \\\hline
\end{tabular}
\label{simulation parameters}
\caption{Simulation parameters}
\end{table}

\subsection{Varying density}

The default value value for the density is $10 / (100^2\cdot 3.14)$.
In the first experiment set, we test the algorithm with the following densities:
$7.5 / (100^2\cdot 3.14)$, $10 / (100^2\cdot 3.14)$, $20 / (100^2\cdot 3.14)$ and $40 / (100^2\cdot 3.14)$, 
Which means that the expected number of neighbours in the communication graph is
about $7.5$, $10$, $20$ and $40$ respectively, since we use the disc graph model for our simulations (c.f. section \ref{sec simuls}), whereas each point of the network can be roughly expected to be sensing covered by $7.5(25/100)^2 = 0.46875$ to $40(25/100)^2 = 2.5$ sensor nodes.
(I.e. we choose density parameters which always guarantee that the network is connected with high probability, however the sensing coverage goes from being partial to dense).

\begin{figure}[hbt]
 \centering
	\subfigure[Distance]{\label{dist1}\includegraphics[width=0.45\textwidth]{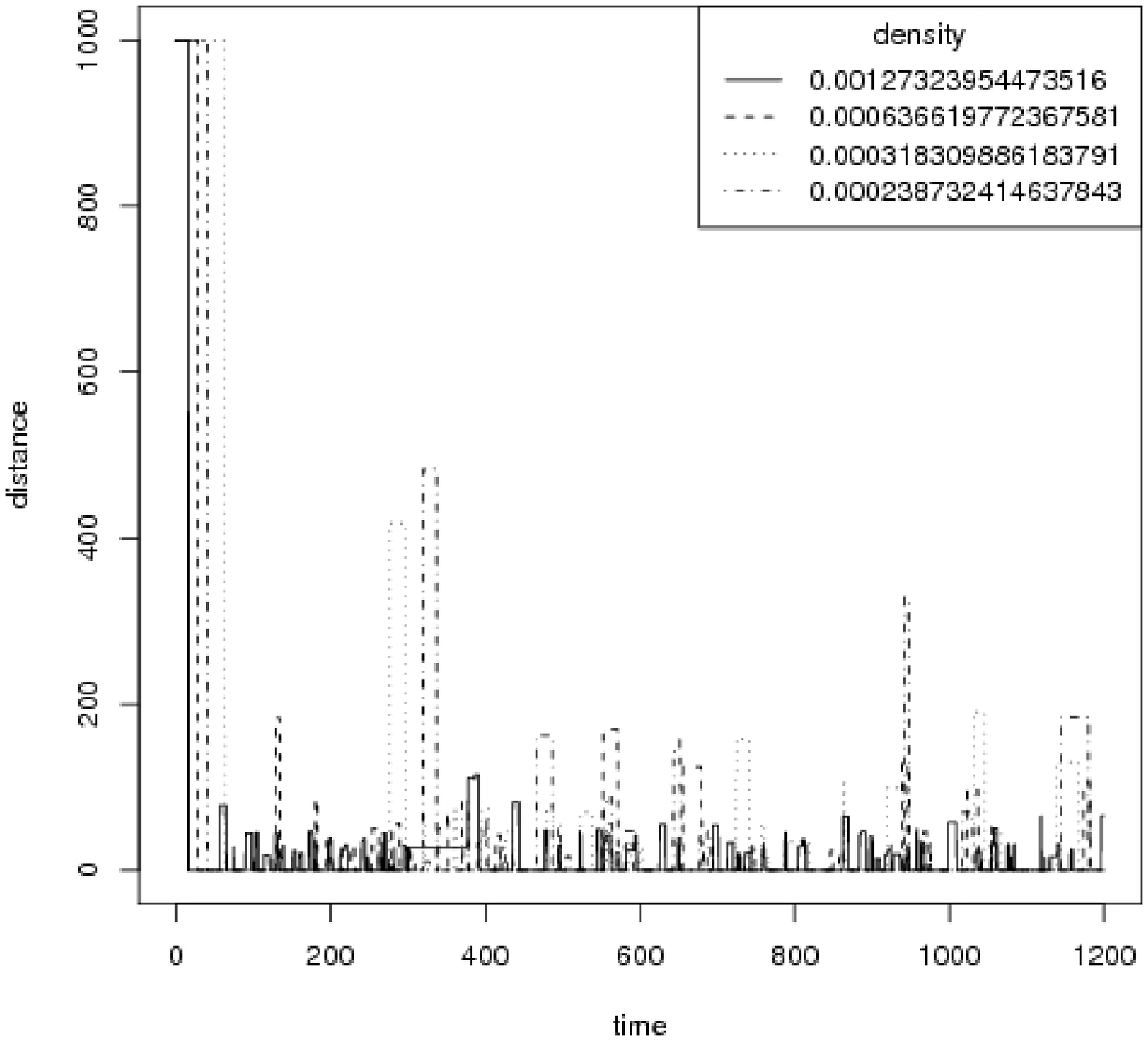}}
	\subfigure[Messages]{\label{messages1}\includegraphics[width=0.45\textwidth]{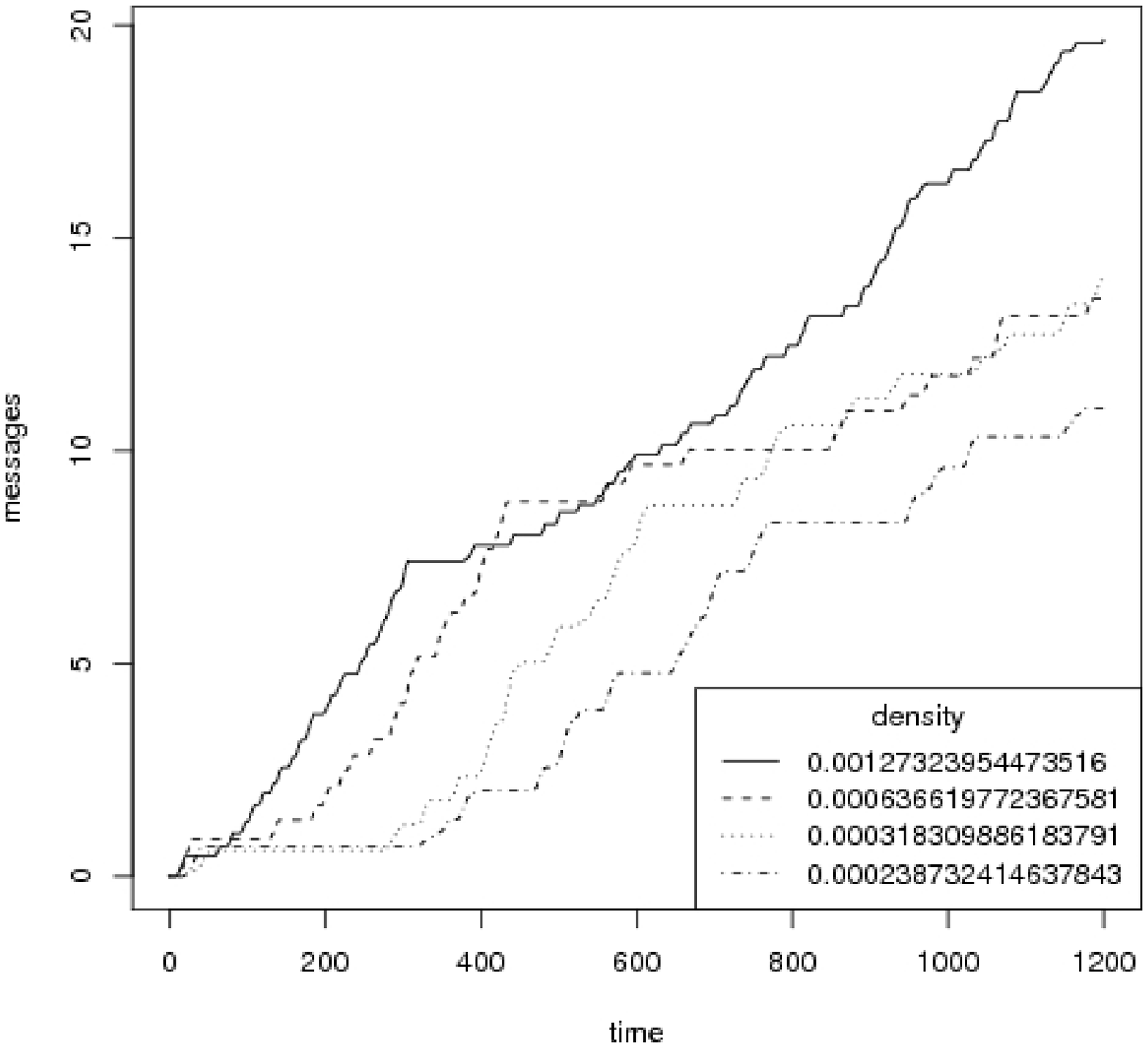}}
\label{fig dens}
\caption{Varying densities}
\end{figure}

The results are shown on figure \ref{fig dens}.

In figure \ref{dist1}, we see that the tracker frequently localizes the target (each time the distance is $0$), whatever the density tested except for the lower density. 
In figure \ref{messages1}, we see that the 
total number of messages per node is very reasonable for all densities. It is also inter sting to notice that the number of spreading messages seems to be correlated with density.
The correlation is not extremely strong (in reticular above a possible threshold density), because other random factors are quite important too
(like the motion pa tern of the target). However, we do observe that the number of
messages diminishes with the density. This is due to the way the heuristic spanning tree is constructed and is a desirable feature (specifically the way one of the nodes always tries to go the further possible from the repulsion point, c.f. section \ref{trace spreading}). 

\subsection{Varying Speed}

\begin{figure}[hbt]
 \centering
	\subfigure[Distance]{\includegraphics[width=0.45\textwidth]{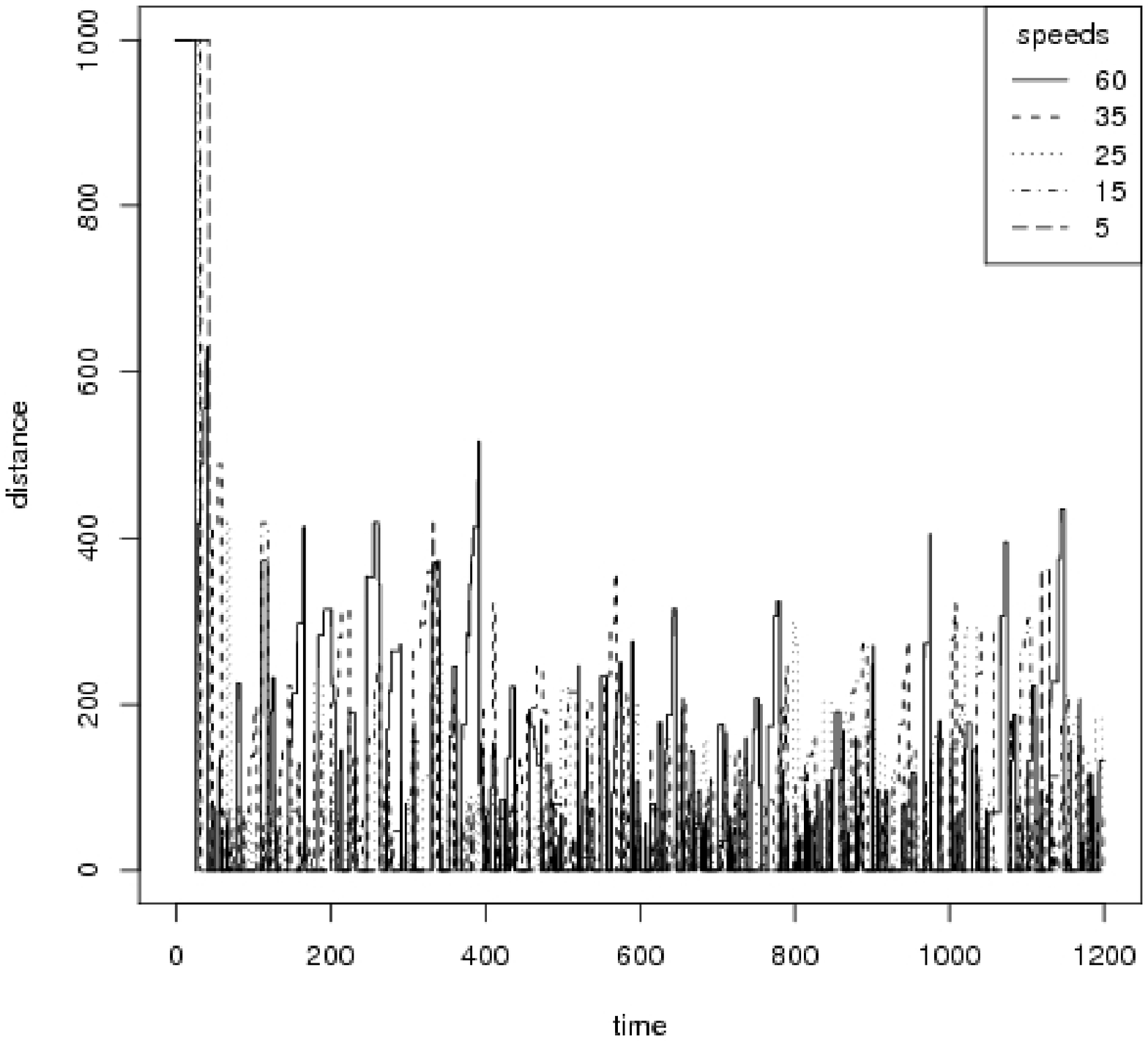}}
	\subfigure[Messages]{\label{messages2}\includegraphics[width=0.45\textwidth]{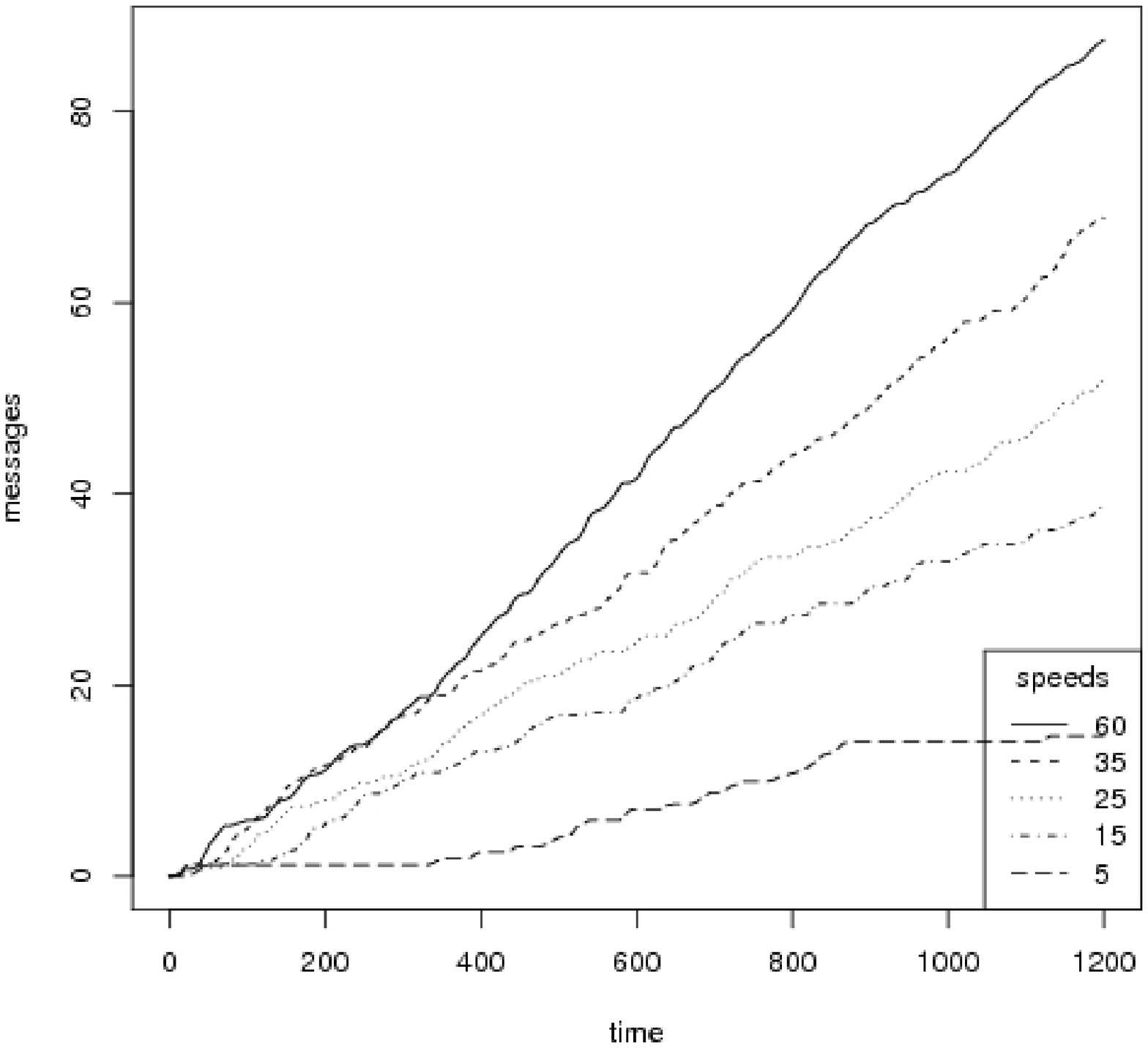}}
\label{fig speed}
\caption{Varying speed}
\end{figure}

The default value value for the target speed is $6$ km/h.
In the second experiment set, we test the algorithm with the following speeds:
$5$, $15$, $25$, $35$ km/h. 
The results are shown on figure \ref{fig speed}.

Again, we notice that for all the tested target speeds, the target is often localised by the tracker. This is quite nice, since it could have been feared that for the highest target speeds the tracker would never have localized the target.
 
Also, the number of messages is kept low. When the target's speed increases, more
detection occurs and thus more traces need to be spread, which explains the increased number of messages with increased speed from figure \ref{messages2}.

\subsection{Varying Sensing Radius}

\begin{figure}[hbt]
 \centering
	\subfigure[Distance]{\includegraphics[width=0.45\textwidth]{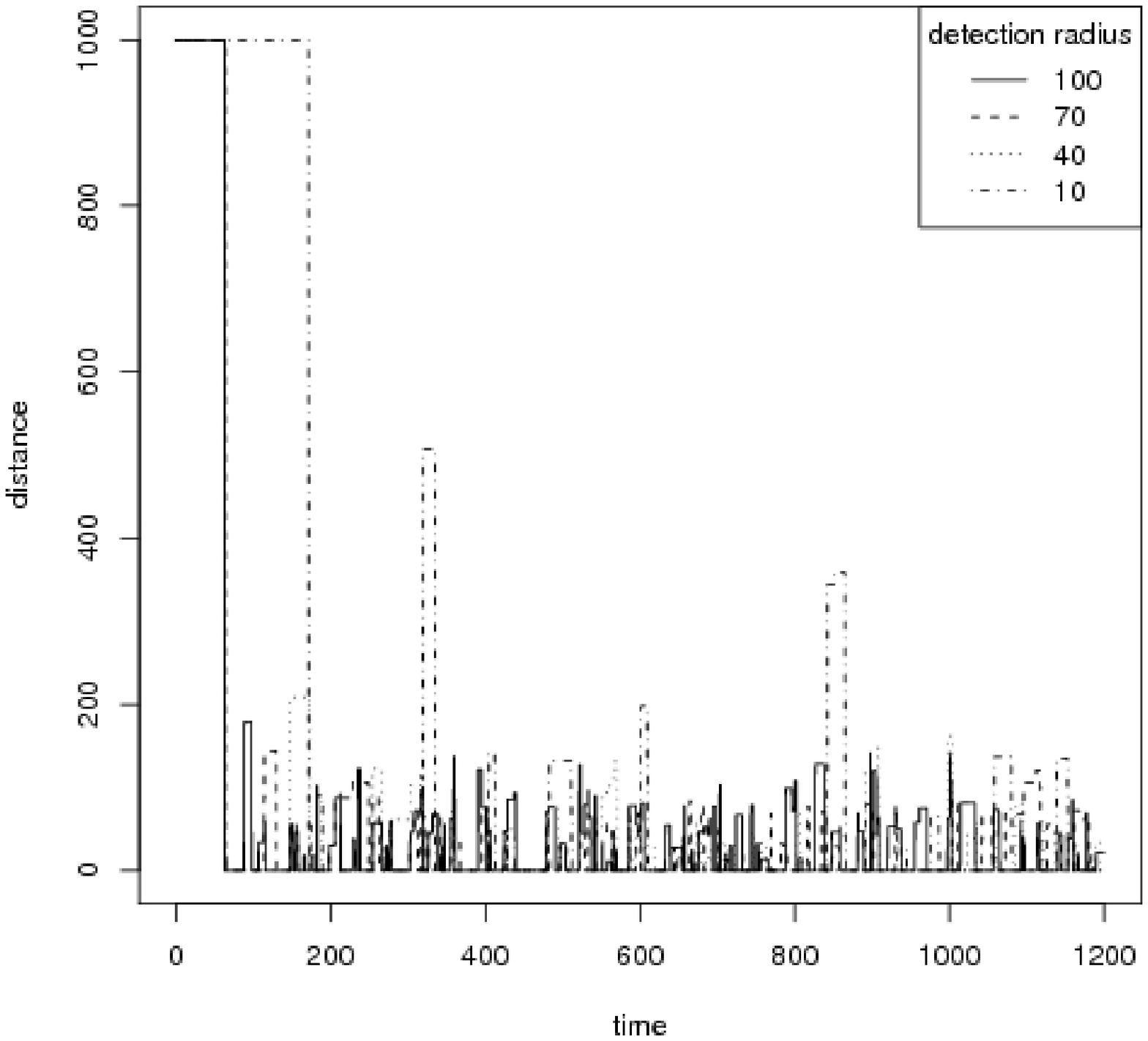}}
	\subfigure[Messages]{\includegraphics[width=0.45\textwidth]{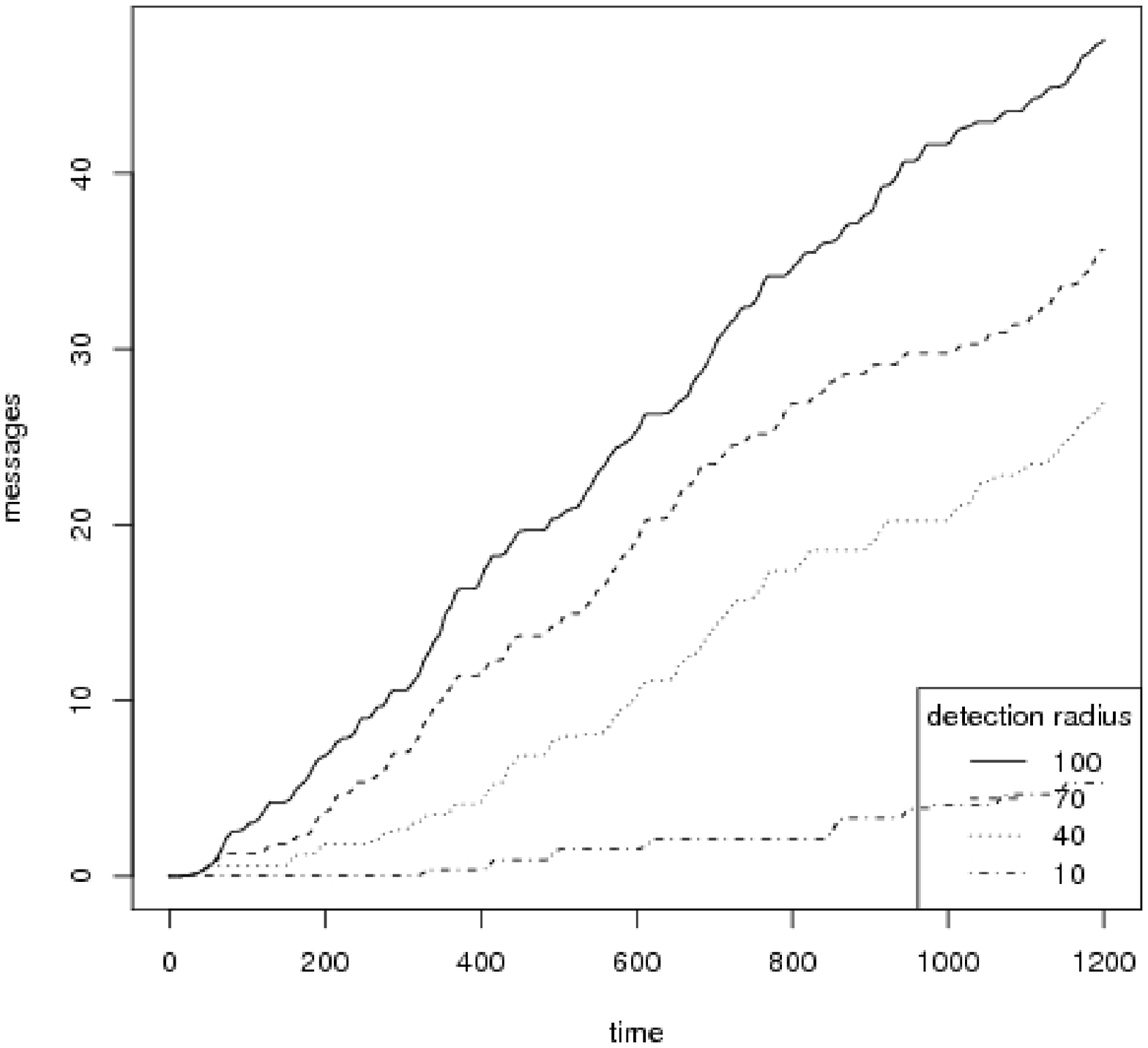}}
\label{fig sens}
\caption{Varying sensing radius}
\end{figure}

The default value value for the sensing radius is $25$ m.
In the third experiment set, we test the algorithm with the following sensing radii:
$10$, $40$, $70$, $100$. 
The results are shown on figure \ref{fig sens}.

We observe the same kind of favorable behavior. 
The target is being exactly localised pretty often, for all sensing radii.
For increasing sensing radius there are more detections by the sensors and thus more messages are being generated, but in every case the number of messages is kept very reasonable.

\subsection{Varying Message Frequency}

\begin{figure}[hbt]
 \centering
	\subfigure[Distance]{\includegraphics[width=0.45\textwidth]{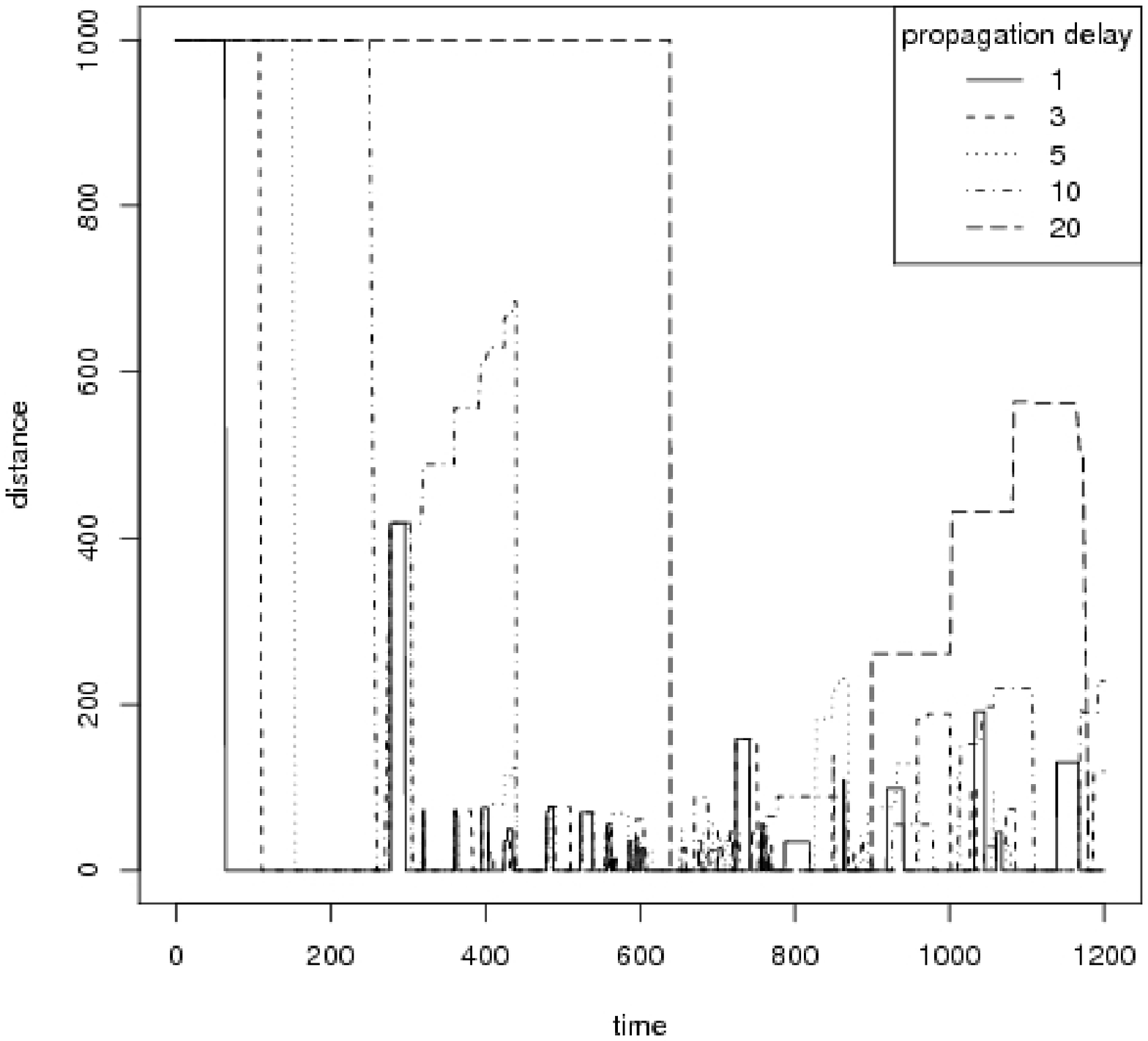}}
	\subfigure[Messages]{\includegraphics[width=0.45\textwidth]{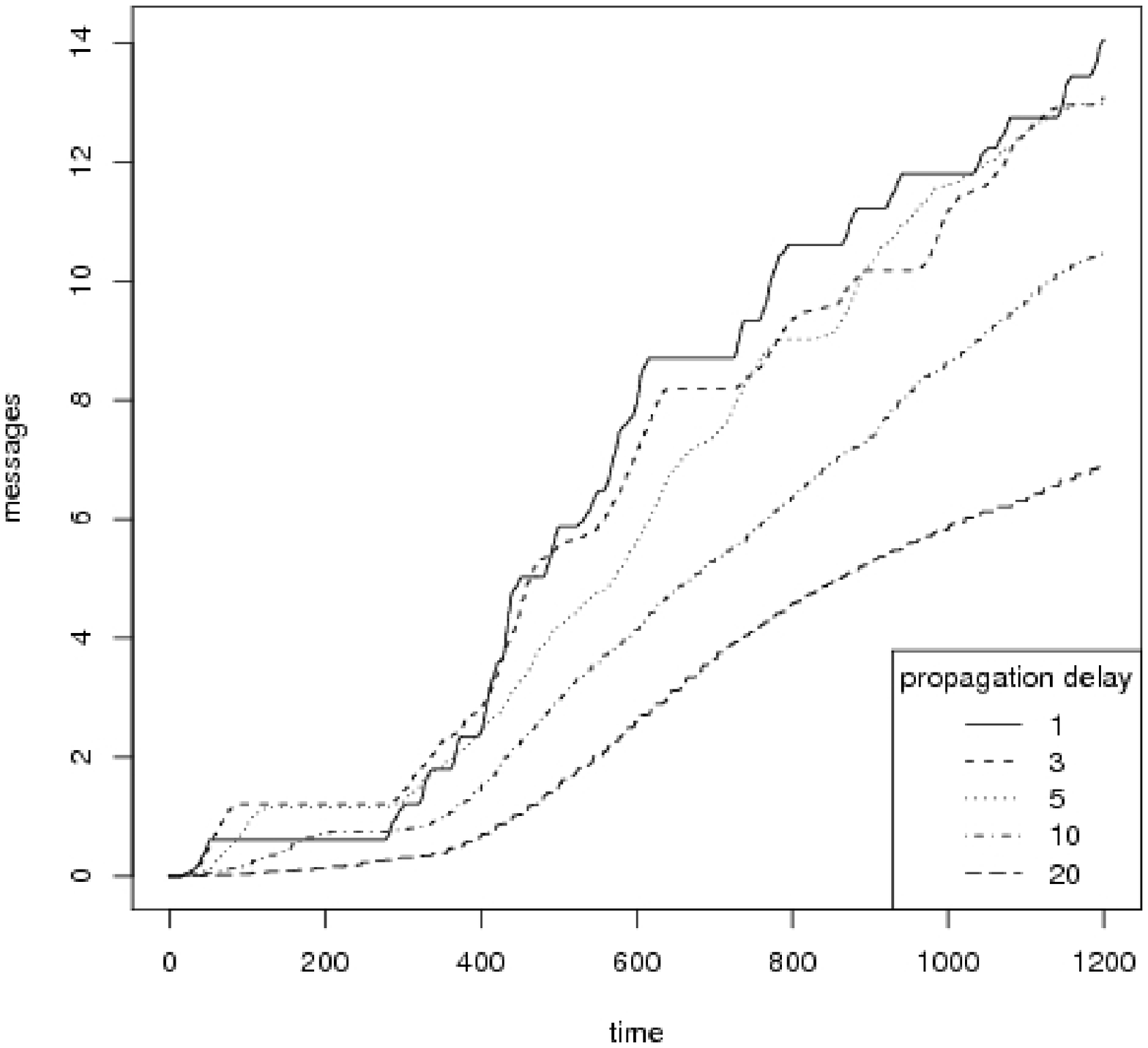}}
\label{fig freq}
\caption{Varying Message Frequencies}
\end{figure}

The default value value for the message frequency is $1$ second,
i.e. the tracking agent moves every second and traces are spread every second.
In the third experiment set, we test the algorithm with the following message propagation frequencies:
$1$, $3$, $5$, $10$, $20$. 
The results are shown on figure \ref{fig freq}.

The most important observation is that, not surprisingly, is that for long propagation delays (e.g. when a message is only sent every $20$ seconds),
it is more difficult to localise the target. 
This is so because with increasing message propagation delay, the tracking agent becomes slower
and is thus less efficient at tracking the target.
Still, except for the longest propagation delays, the target is being localised regularly.
A very nice result is that the number of messages sent increases only slowly with reducing propagation delays. That is, if messages are sent twice more often, the total number of messages is far from doubling. This is due to the inhibition mechanism: when messages are sent quickly, the spreading of a trace is more likely to be inhibited by a previous trace which has not yet vanished with time. On the contrary, when propagation delay are long, the spreading of a new trace is likely to span a larger portion of the network, since previous trace intensities will be low due to the time-intensity dissipation implied by our model.

Therefore, precision can be traded for increased traffic.

\section{Simulation Implementation Details}
\label{sec simuls}

We used a \texttt{ruby} implementation to make our simulations.
In our simulations, we use the unit disc graph model for building a high level abstratcion
of the communication graph associated with the sensor network. The target movement is modelled using the random waypoint model. In this model, the target chooses a random point and moves to
it at the chosen speed (which was a fixed adjustable parameter in our simulations).
As a consequence, the target moves along line segments. If the target is moving from point
$A$ to point $B$, starting at time $t_0$ and arriving at time $t_1$, we used the following
model for target detection by the sensor nodes: 
if a sensor node $n$ is at distance $d(\overline{AB},n)$ of the segment $\overline{AB}$
and that $d(\overline{AB},n)$ is smaller than than the sensing radius $d_{dtx}$ (c.f. table \ref{simulation parameters}), then the node detects the presence of the target at time $t$, 
where $t = \frac{d(x,A)}{d(A,B)}(t_1-t_0)$, where $x$ is the only point of the segment 
$\overline{AB}$ such that $d(x,n) = d(\overline{AB},n)$. This means that an event can occur
at times taking value in the real numbers (or at least their floating point representation in the programming language). We feel that this fairly complicated simulation environment, when
compared to an approach that would discretize time in rounds, is much more precise.
At the cost of complicating the implementation of the simulation platform and putting higher computation loads on its execution, we thus used a fairly detailed simulation environment. Although events occur at real number valued times, there is only a finite number of event during each time interval. We use this
to store events on an \textsl{scheduled events stack} ordered by time, and execute the events starting at the top of the stack. For example, when the target starts moving from point $A$ to $B$, we first compute all detections by sensor nodes and put them on the scheduled events stack, ordered by time.
We then ``execute'' the first detection by picking the scheduled event at the top of the stack. The top event, assuming it is an target detection by a sensor node, implies initializing a trace on the detecting node. Because this trace will have to be spread a few seconds later (depending on the simulation parameters), we create a ``spreading'' event and insert it on the scheduled event stack and sort the stack according to the scheduled execution time of the event (a full sorting is not required, since the stack is always kept sorted, it in fact suffices to introduce the new event at the proper place in the stack). Summarizing, the executing flow of the simulator is (1) Pick the event at the top of the stack, (2) Execute the event (e.g initialize a trace) (3) Determine all future events implied by the execution and insert them on the scheduled events stack. Start at point (1) again.
\section{Conclusions} 
We have proposed a lightweight target tracking protocol for wireless sensor networks.
inspired by an analogy with the way a lion tracks down an antelope in the Savannah by following it's smell.
In our protocol, targets leave traces behind them.
The intensity of those traces (like the odor of the antelope) decreases with time.
The sensor network propagates those traces (like the wind propagates the smell of the antelope) using an innovative propagating process aiming at spanning the network with a tree
of degree 2 with non overlapping branches.
The protocol was evaluate trough extensive simulations under representative network operation
regimes. It was shown that the protocol is successful at letting a tracking agent localize the target regularly, and that the message overhead was kept low. More precisely, it was shown that the protocol scales particularly well in terms of network density since the localization process is efficient and the message overhead is low, for all tested densities.
We also show that the network is capable of tracking targets moving at different speeds (from slow to quite fast). We have also tested different sensing coverage regimes (from not well covered to redundantly covered), and showed that although traffic increases with increased sensing range, target localisation is achieved for all regimes with reasonable message overhead. Finally, an important finding is that diminishing propagation delays (e.g. by increasing the duty cycle of sensor nodes) does not augment too much the message overhead because of the trace spreading inhibition mechanism.


\begin{thebibliography}{20}

\bibitem{intro3} I. F. Akyildiz, W. Su, Y. Sankarasubramaniam, and E. Cayirci. Wireless sensor networks: a survey. Computer Networks, 371,
38:393$�$422, March 2002.

\bibitem{grid}
K. Chakrabarty, S. S. Iyengar, H. Qi, and E. Cho, Grid coverage for surveillance and
target location in distributed sensor networks, IEEE Trans. Comput. 51(12) (2002).



\bibitem{das}
R. Gupta and S. R. Das, Tracking moving targets in a smart sensor network, Proc VTC
Symp., Volume: 5, pp. 3035 - 3039, 2003.

\bibitem{liu}
J. Liu, J. Liu, J. Reich, P. Cheung, and F. Zhao, Distributed group management for track
initiation and maintenance in target localization applications, Proc. Int. Workshop on
Information Processing in Sensor Networks (IPSN), 2003.





\bibitem{nikole} S. Nikoletseas and P. Spirakis, 
Efficient Sensor Network Design for Continuous Monitoring of Moving Objects,
CTI Technical Report, 2007.


\bibitem{book} Rajeev Shorey, A. Ananda, Mun Choon Chan, Wei Tsang Ooi,
Mobile, Wireless, and Sensor Networks: Technology, Applications, and
Future Directions, Wiley, 2006.

\bibitem{tseng} Yu-Chee Tseng, Sheng-Po Kuo, Hung-Wei Lee and Chi-Fu Huang,
Location Tracking in a Wireless Sensor Network by Mobile Agents and Its Data Fusion Strategies,
in Proc. of IPSN 2003,pp. 625 - 641, 2003.


\bibitem{intro2} B.Warneke, M. Last, B. Liebowitz, and K.S.J. Pister, Smart dust: communicating with a cubic-millimeter computer, Computer, 369
34:44$�$51, January 2001.

\bibitem{infocom} W. Zhang and G. Cao, Optimizing tree reconfiguration for mobile target tracking in
sensor networks, Proc. IEEE InfoCom, 2004.

\bibitem{guibas} Feng Zhao, Leonidas Guibas,
Wireless Sensor Networks: An Information Processing Approach, 
Publisher: Morgan Kaufmann, 2004.


\end{thebibliography}
\end{document}